\documentstyle[psfig,epsf,conf-X]{article}
\begin{document} 
\small
\def\cgs{ ${\rm erg~cm}^{-2}~{\rm s}^{-1}$ } 
\heading{%
%
The BeppoSAX High Energy Large Area Survey (HELLAS): a progress report
}
\par\medskip\noindent
\author{%
A. Comastri$^1$, F. Fiore$^{2,3,4}$, C. Vignali$^{1,5}$, F. La Franca$^6$,
G. Matt$^6$
}
\address{%
Osservatorio Astronomico di Bologna, via Ranzani 1, I--40127
Bologna, Italy 
}
\address{%
Osservatorio Astronomico di Roma, Via Frascati 33,
I--00044 Monteporzio, Italy
}
\address{%
BeppoSAX Science Data Center, Via Corcolle 19, I--00131 Roma, 
Italy
}
\address{%
Center of Astrophysics, 60 Garden Street, 
Cambridge MA 02138 USA
}
\address{%
Dipartimento di Astronomia, via Ranzani 1, I--40127
Bologna, Italy
}
\address{%
Dipartimento di Fisica, Universit\`a degli Studi ``Roma Tre",
Via della Vasca Navale 84, I--00146 Roma, Italy 
}

\begin{abstract}

The integrated emission of highly obscured AGN is expected to provide a 
major contribution to the X--ray energy density in the Universe: 
the X--ray background (XRB).
The study  of these objects is possible only at energies where the
effects of absorption are less severe.
For this reason we have carried out the {\sf BeppoSAX} High Energy LLarge 
Area Survey 
in the hardest band (5--10 keV) accessible so far with imaging X--ray 
instruments.
The source surface density at the survey limiting flux accounts
for a significant fraction (20--30 \%) of the hard XRB.
The X--ray data complemented by multiwavelength follow--up observations 
suggest that a large fraction of the hard sources are AGN and that 
X--ray absorption with column densities in the range 
10$^{22-23.5}$ cm$^{-2}$ is common among them.
The great diversity in their optical--near--IR properties 
suggests that the optical appearance of obscured sources 
is a function of the X--ray luminosity. 
We briefly discuss the implications of these findings for the XRB models.
\end{abstract}
\section{Introduction}

The identification breakdown of soft X--ray surveys has clearly
established that AGN are by far the dominant population of the
X--ray sky (see \cite{L99} for the most recent 
update) implying that about 60--70 \% of the soft XRB  
 has been already resolved into AGN at a limiting 
0.5--2 keV flux of $\sim$ 10$^{-15}$ \cgs. 
Thanks to the imaging capabilities of {\sf ASCA} and {\sf BeppoSAX} detectors
the number of hard X--ray selected optically identified objects is 
increasing and several dozens of identifications
are now available confirming that most of these sources 
are indeed AGN (\cite{F99}, \cite{Ak00}).
Their contribution to the 2--10 keV XRB 
is of the  order of 20--25 \% at a flux limit 
of $\sim$ 5 $\times$ 10$^{-14}$ \cgs.

The present findings provide support to the XRB synthesis models
which, in the framework of the unified scheme,
assume that a mixture of unabsorbed and absorbed AGN 
can account for almost the entire XRB spectral intensity in the
2--100 keV energy range (\cite{SW89}, \cite{C95}, \cite{Gi99}, \cite{PLM99}) 
Even though the so far proposed models differ in several 
of the assumptions concerning the AGN luminosity function and the 
X--ray spectral properties, they all agree in predicting 
that the fraction of obscured AGN is rapidly increasing 
towards high energy and faint fluxes.
It turns out that these sources can be efficiently discovered 
with sensitive hard X--ray surveys.

In order to quantitatively test AGN synthesis models for the XRB  
we have carried out an X--ray  survey in the hardest band
accessible with the present imaging detectors: 
the 5--10 keV {\sf BeppoSAX} High Energy LLarge Area Survey (HELLAS).

\section{Source Counts}

About 80 square degrees of sky have been surveyed in the 5--10 keV
band using several {\sf BeppoSAX} MECS (\cite {B97}) high Galactic
latitude ($|b|>20$ deg) fields. All MECS pointings cover different
sky positions.
The fields were selected among public data (as March 1999) and our 
proprietary data excluding those fields
centered on extended sources and bright Galactic objects.
A robust detection algorithm has been used on the coadded MECS1
MECS2 and MECS3 (or MECS2 plus MECS3 after the failure of MECS1
in May 1997) images and the quality of the 
detection has been checked interactively for all the 
147 sources of the final sample.
Background subtracted count rates were converted to 5--10 keV fluxes
assuming a power law spectrum with energy index $\alpha_E$ = 0.6
A detailed description of the survey and the detection procedure 
is presented in \cite {F00}.

The 5--10 keV logN--logS of the HELLAS sources is reported in Figure 1. 
The source surface density 
at the survey limit (4.8 $\times10^{-14}$ \cgs) is  17$\pm$6 sources 
per square degree.
The error bars on the binned integral counts account for both
the statistic and systematic uncertainties, the latter due to 
the lack of information on the intrinsic spectrum of the faint sources. 
The systematic error  has been 
estimated assuming a range of spectral shapes (0.2 $<$ $\alpha_E$ $<$ 1.0) 
to convert count rates into fluxes.
\begin{figure}[t]
%
\centerline{
\psfig{figure=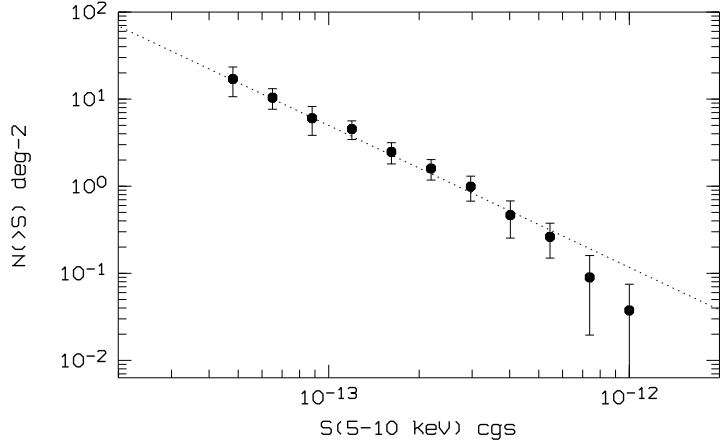,width=9cm,bbllx=30pt,bblly=30pt,bburx=730pt,bbury=545pt}}
\caption[]{The 5--10 keV HELLAS counts compared with the 
{\sf ASCA} 2--10 keV counts (dashed line). A power law spectrum with
$\alpha_E$ = 0.6 has been adopted for the flux conversion.}
\end{figure}
The HELLAS counts are in very good agreement with the recent {\sf ASCA} results
(\cite{C98}, \cite{U99}) obtained in the 2--10 keV band (Figure 1). 
Given that both the 
HELLAS and the {\sf ASCA} counts have been computed assuming the same 
prescription for the spectral slope the comparison is straightforward.
The dashed line in Figure 1 corresponds to the best--fit to the {\sf ASCA} 
counts computed by \cite {DC99} and converted in the 5--10 keV band
with $\alpha_E$ = 0.6.
The cumulative flux of the HELLAS sources 
accounts for a significant fraction (20--30 \%) of the 5--10 keV XRB.
The main uncertainty on the resolved fraction is due to the
still poorly understood normalization of the extragalactic XRB spectrum
as measured by the different satellites (see \cite{C00} and
\cite{Ve99} for more details).

\section{Hardness ratios}

The HELLAS sources are too faint to perform a spectral fit.
In order to study their spectral properties 
we have computed for each
source, whenever possible, two X--ray colours defined as 
HR1 = (M$-$L)/(M+L) and HR2 = (H$-$M)/(H+M)
where H, M and L are the number of counts in the 
H=4.5--10 keV  M=2.5--4.5 keV and L=1.3--2.5 keV energy ranges.
A wide range of spectral properties is evident from the analysis of
the color--color diagram reported in 
Figure 2. For example extremely hard sources, with nearly all the photons 
detected only above 4.5 keV, populate the upper right part of the diagram.
\begin{figure}[h]
\centerline{
\psfig{figure=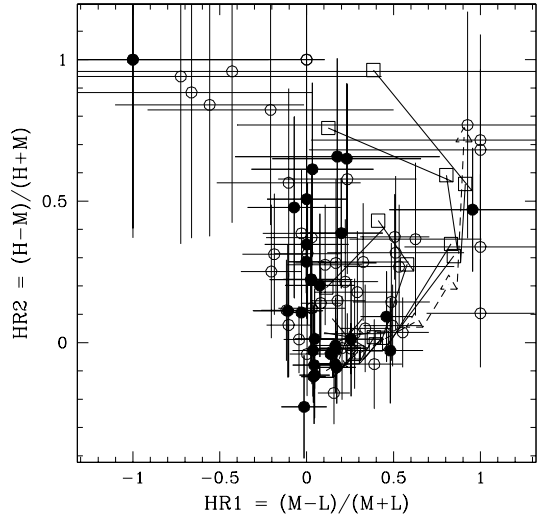,width=8cm}}
\caption[]{The X--ray color--color diagram for the 58 sources
detected above 3.8 $\sigma$ (see text for details). The optically identified 
objects are reported with filled symbols.}
\end{figure}
The X--ray colours depend from the intrinsic spectrum and 
from the source redshift. The colours expected
for a range of spectral shapes at different redshifts has been computed
following the same procedure adopted by \cite{DC991} for their {\sf ASCA} 
survey.

The big star at HR1 = 0.25 and HR2 = 0 represents an unabsorbed power law
spectrum with $\alpha_E$ = 0.4.
The rightmost solid curve connecting the open squares indicates the 
colours for an AGN--like power law ($\alpha_E$ = 0.8) 
absorbed by an increasing value of the column density : log $N_H$ =  
0, 22, 22.7, 23, 23.7, 24 at z=0. The dashed curve is the same but at z=0.4.
The innermost solid curves have been computed at z=0 assuming the same
model but allowing a fraction of 10\% and 1\% respectively to be unabsorbed.
These models should be considered as indicative and indeed they do not 
cover the entire diagram. As first noted by \cite{DC991}, 
more complicate spectral shapes, 
such as those characterizing the sources in the upper left portion 
of the plot, might be present.
{\sf ASCA} follow--up pointed observations of two relatively bright
HELLAS sources \cite{VS00}  allowed to collect enough counts 
to perform a more detailed spectral analysis.
In both of the cases the data are consistent with a hard absorbed 
power law spectrum making us confident on the robustness
of the hardness ratio results. 

\section{Optical Identifications}

The cross--correlation of the HELLAS sample with various source catalogs
provided 25 coincidences (19 AGN: 7 radio--loud, 12 radio quiet 
and 6 clusters of galaxies). In addition, optical spectroscopic 
follow--up observations have been performed and 22 new identifications
(18 AGN) are available (\cite{F99}, \cite{FLF00}). 
A detailed discussion of the optical identifications 
is beyond the purposes
of this paper and will be presented by \cite{FLF00}.

The average X--ray spectrum as inferred form the softness ratio
value S$-$H/S+H (where S is the number of counts in the 1.3--4.5 keV
energy range) is plotted in Figure 3 versus the source redshift 
and optical classification.  

The most important results are the following:

$\bullet$ The fraction of type 2 objects (including in this
class Seyfert types 1.8--1.9--2.0 and quasars with a red optical
continuum) is of the order of 40--45 \%.
This percentage is higher than in other optically identified samples
of X--ray selected sources from {\sf ROSAT} and {\sf ASCA} surveys.

$\bullet$ The degree of obscuration as inferred from the hardness ratio
analysis described in the previous section 
seems to be uncorrelated with the optical reddening indicators, 
such as the optical line widths and line ratios.\\
\begin{figure}[h]
\centerline{
\psfig{file=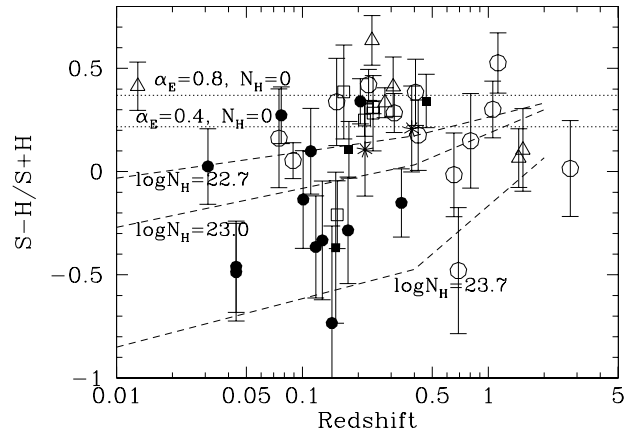,width=9.0cm,bbllx=0pt,bblly=30pt,bburx=730pt,bbury=545pt}}
\caption[]{The softness ratio versus redshift for the so far identified
sources. Different symbols mark different sources: open circles = broad line
quasars with a blue optical continuum, stars = broad lined quasars
with a red optical continuum, filled circles = AGN of types 1.8--1.9--2.0,
filled squares = emission line galaxies , open triangles = radio--loud AGN
open squares = clusters of galaxies. The two dotted lines represent the
expected softness ratio for a power law with $\alpha_E$=0.4 and
$\alpha_E$=0.8. Dashed lines show the softness ratios of absorbed power 
law models (for $\alpha_E$ = 0.8 and log$N_H$ = 22.7, 23, 23.7, from top 
to bottom) with the absorber at the source redshift.}
\end{figure}

$\bullet$ The softness ratio of a few high luminosity
broad lined quasars with blue optical colours implies 
X--ray absorption by a substantial column density (log $N_H >$ 23).

$\bullet$ Optical and near--infrared photometry of ten HELLAS 
sources carried out at the Italian National Telescope Galileo 
(TNG) indicates that the optical colors of type 1.8--2.0 and red
AGN are dominated by the host galaxy, though the obscured AGN 
contributes to some of the infrared emission (\cite{M00}, \cite{V00}, 
\cite{VS00}). 

\section{A quick comparison with XRB synthesis models}

The observed 5--10 keV logN--logS is compared with the AGN number counts 
predictions \cite{C95} in the same energy range (Figure 4).
Given that in the HELLAS band a column as high as log $N_H$ = 23
is needed to significantly reduce the photon flux, 
the contribution of sources with a different degree of obscuration 
is reported splitted into 3 classes : relatively unobscured 
log $N_H <$ 23, obscured Compton thin 23 $<$ log$N_H <$ 24 and
almost Compton thick log $N_H >$ 24.
According to the model predictions relatively unobscured sources outnumber 
absorbed objects.
The expected fraction of Compton thin 
sources ranges from 25 to 35 \% while the number of Compton thick 
sources is always negligible.
Given that about one third of the AGN reported in Figure 3
have a softness ratio value consistent with a column 
density 23 $<$ log$N_H <$ 24 and none is Compton thick 
the agreement with the model predictions is remarkable.
These findings are at variance with the relatively high 
space density of Compton thick AGN in the local Universe 
\cite{M98}, \cite{R99}. We note, however, that the fraction of Compton thick
objects has been estimated only for optically selected, nearby 
Seyferts and thus might not be representative of the more distant 
X--ray selected population.  

\begin{figure}
\centerline{
\psfig{figure=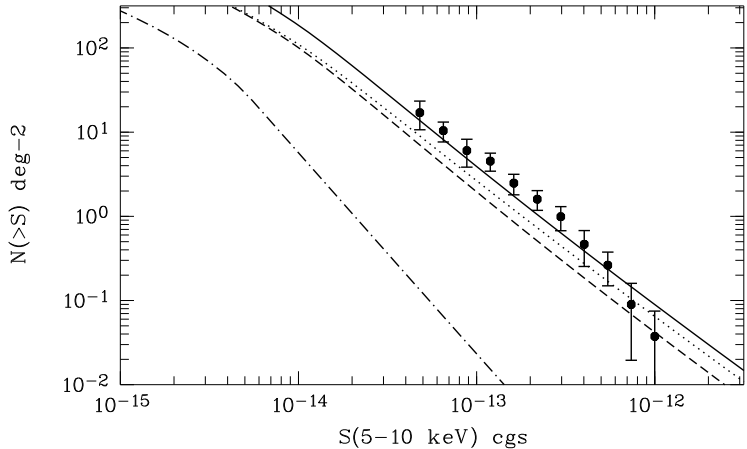,width=8.4cm,bbllx=100pt,bblly=100pt,bburx=730pt,bbury=545pt}}
\caption[]{The 5--10 keV counts compared with the prediction
of XRB synthesis models (\cite{C00}). The solid line represents
the total AGN contribution, the other lines the number counts of AGN with
various degrees of intrinsic obscuration: log$N_H <$ 23 (dotted),
 23 $<$ log$N_H < $ 24 (dashed), log$N_H >$ 24 (dot--dashed)}
\end{figure}

\section{Discussion}

The hard X--ray sky, as surveyed by {\sf BeppoSAX}, is dominated by 
AGN with a wide range of X--ray and optical properties.
At first glance the present data are in line with what expected
from the standard synthesis models for the XRB \cite{C95}. In addition, 
new trends characterizing the AGN population are emerging.  

The large dispersion in the hardness/softness ratios suggests 
that the spectral properties of the HELLAS sources
cannot be explained only with a distribution of absorbing column densities
as assumed in the XRB models. In agreement with {\sf ASCA} findings  
\cite{DC991}, the presence of additional soft 
X--ray emission above the hard absorbed component is required.
The ``soft excess" is likely to be due either to a fraction of 
the nuclear component scattered in the line of sight 
or to an incomplete covering of the central radiation or 
to thermal emission originating in a starburst region and/or due
to X--ray binaries and supernova remnants.
The ``soft excess" intensity, which ranges from a few percent
to 10--20 \% of the total X--ray flux, is not relevant
as far as the fit to the hard XRB is concerned, 
however it should be consistently taken into account 
when model predictions are compared to hard and soft X--ray counts.
Indeed, if soft excesses are common among obscured AGN, as the HELLAS results
indicate, these sources might well be detectable 
in the soft X--ray band even in the presence of 
substantial obscuration simply because the flux limit of the 
{\sf ROSAT} soft X--ray surveys is about a factor 
30--40 deeper than the limit actually reached in the hard X--ray band.

Another interesting and unexpected result concerns the optical and
near--infrared properties of the hard X--ray selected sources.
Even though detailed studies have been 
carried out only for a small HELLAS subsample the 
results do clearly indicate that:

$\bullet$ The broad band optical (U,B,V,R,I) and near--infrared
(J, K) colors of HELLAS AGN, spectroscopically classified as type 1.8--2.0
Seyferts or red quasars, are undistinguishable from those of normal 
passive galaxies. 

$\bullet$ There is an increasing evidence that the X--ray absorption 
properties and the optical appearance of AGN change with redshift and/or 
luminosity, suggesting that high luminosity, highly obscured
quasars are present among optically broad lined blue objects.

If confirmed by future observations, these findings would imply that
the X--ray obscured AGN responsible for a large fraction 
of the hard XRB energy density could be ``hidden" among 
objects which would be optically classified 
either as normal galaxies or as ``normal" blue quasars.
It is also worth noting that the redshift of obscured 
AGN could be in principle obtained by deep photometric observations
of their host galaxies.

A step forward in the study of the sources of the hard XRB 
will be achieved by the foreseen {\sf XMM} and {\sf Chandra} surveys.  
In particular the X--ray data will allow to obtain an 
unbiased estimate of the absorption distribution (a key parameter
of the XRB models), while optical and near--infrared follow--up 
observations are likely to provide new insights on the 
nature of the absorbing gas (such as the dust--to--gas ratio) 
and on the morphology of the host galaxies of obscured AGN.

As a final remark we note that, while type 2 quasars are probably numerous,
it is likely that they have been elusive so far because 
have been searched for in the optical rather than in X--rays. In most 
luminous and/or distant absorbed quasars the Narrow Lines Region may 
be also obscured or lacking altogether 
(NGC~6240 is a classical example \cite{Pa99}), 
and therefore they are missed in optical spectroscopic surveys. Deep
{\sf Chandra} and {\sf XMM} surveys will hopefully settle this long--standing
issue, and then provide a key test for XRB synthesis models.

\acknowledgements{
We thank the {\sf BeppoSAX} SDC, SOC and OCC teams for the successful
operation of the satellite and preliminary data reduction and
screaning, P. Giommi, R. Maiolino, L.A. Antonelli, S. Molendi, M. Mignoli, 
R. Gilli, G. Risaliti (the ``HELLAS boys")
for the fruitful collaboration, M. Salvati, G.C. Perola and G. Zamorani
for useful discussions. Partial support from ASI contract ARS--99--75 and 
MURST grant Cofin98--02--32 is acknowledged. 
}

\vfill

\begin{iapbib}{99}{
\bibitem{Ak00} Akiyama M., et al., 2000, \apj in press
\bibitem{B97} Boella G., et al., 1997, A\&AS 122, 299 
\bibitem{C98} Cagnoni I., Della Ceca R., Maccacaro T., 1998, \apj 493, 54  
\bibitem{C95} Comastri A., et al., 1995, \aeta 296, 1 
\bibitem{C00} Comastri A., 2000, Astr. Lett \& Comm., submitted  
\bibitem{DC991} Della Ceca R., et al., 1999, \apj 524, 674     
\bibitem{DC99} Della Ceca R., et al., 2000, astro-ph/9912016 
\bibitem{F99} Fiore F., et al.,  1999, \mn 306, L55
\bibitem{F00} Fiore F., Vignali C., Comastri A., et al., 2000, \mn submitted  
\bibitem{Gi99} Gilli R., Risaliti G., Salvati M., 1999, \aeta 347, 424  
\bibitem{FLF00} La Franca F., et al., 2000, in preparation 
\bibitem{L99} Lehmann I., et al., 1999, \aeta in press (astro-ph/9911484)
\bibitem{M98} Maiolino R., et al., 1998, \aeta 338, 781  
\bibitem{M00} Maiolino R., et al., 2000, \aeta submitted
\bibitem{PLM99} Pompilio F., La Franca F., Matt G., 2000, \aeta 353, 440 
\bibitem{R99} Risaliti G., Maiolino R., Salvati M., 1999, \apj 522, 157 
\bibitem{SW89} Setti G., Woltjer L., 1989, \aeta 224 L21  
\bibitem{U99} Ueda Y., et al., 1999, \apj 524, L11
\bibitem{Ve99} Vecchi A., Molendi S., Guainazzi M., et al., 1999, \aeta 349, L73 
\bibitem{VS00} Vignali C., Comastri A., Fiore F., 2000, these proceedings
\bibitem{V00} Vignali C., et al., 2000, MNRAS in press   
\bibitem{Pa99} Vignati P., et al., 1999, \aeta 349, L57 
}
\end{iapbib}
\end{document}